\documentstyle[11pt,aaspp4]{article}  
\def\etal{{\it et~al.~}}
\def\bsax{{\it BeppoSAX~}}
\def\ginga{{\it Ginga~}}
\def\einstein{{\it Einstein~}}

\def\asca{{\it ASCA~}}
\def\rosat{{\it ROSAT~}}
\def\rxte{{\it RXTE~}}

\def\erg{$~erg~cm^{-2}~s^{-1}~$}

\begin{document}

\newcommand{\lessim}{\ \raise -2.truept\hbox{\rlap{\hbox{$\sim$}}\raise5.truept
	\hbox{$<$}\ }}			

\title{Hard X-ray emission from the galaxy cluster A2256}

\author{Roberto Fusco-Femiano} 
\affil{Istituto di Astrofisica Spaziale, C.N.R., via del Fosso del Cavaliere, 
I-00133 Roma, Italy - dario@saturn.ias.rm.cnr.it}
\author{Daniele Dal Fiume}
\affil{TESRE, C.N.R., via Gobetti 101, I-40129 Bologna, Italy}
\author{Sabrina De Grandi}
\affil{Osservatorio Astronomico di Brera, via Bianchi 46, I-23807 Merate, Italy}
\author{Luigina Feretti}
\affil{Istituto di Radioastronomia, C.N.R., via Gobetti 101, I-40129 Bologna, 
Italy}
\author{Gabriele Giovannini}
\affil{Istituto di Radioastronomia, C.N.R., via Gobetti 101, I-40129 Bologna, 
Italy}
\affil{Dip. di Fisica, Univ. Bologna, via B. Pichat 6/2, I-40127 Bologna, Italy}
\author{Paola Grandi}
\affil{Istituto di Astrofisica Spaziale, C.N.R., via del Fosso del Cavaliere,
I-00133 Roma, Italy}
\author{Angela Malizia}
\affil{BeppoSAX SDC, via Corcolle 19, I-00131 Roma, Italy}
\author{Giorgio Matt}
\affil{Dip. Fisica, Univ. Roma Tre, via della Vasca Navale 84, 
I-00146 Roma, Italy}
\author{Silvano Molendi}
\affil{Istituto di Fisica Cosmica, C.N.R., via Bassini 15, I-20133 Milano, 
Italy}

\begin{abstract}

After the positive detection by \bsax of hard X-ray radiation up to 
$\sim$80 keV in the Coma cluster spectrum, we present evidence
for nonthermal emission from A2256 in excess of thermal emission at a 
4.6$\sigma$ confidence level. In addition to this power law component,
a second nonthermal component already detected by \asca 
could be present in the X-ray spectrum of the cluster, not surprisingly
given the complex radio morphology of the cluster central region.
 The spectral index of the hard tail detected by the PDS onboard \bsax is 
marginally consistent with
that expected by the inverse Compton model. A value of
$\sim 0.05~\mu G$ is derived for the intracluster magnetic field  
of the extended radio emission in the northern regions of the cluster, while  
a higher value of $\sim 0.5~\mu G$ could be present in the central radio halo,
likely related to the hard tail detected by \asca. 
\end{abstract}


\section{ Introduction}
Nonthermal hard X-ray (HXR) radiation has been detected for the first 
time in the Coma cluster by \bsax (Fusco-Femiano \etal 1999) and 
{\it RXTE} (Rephaeli, Gruber \& Blanco 1999), while marginal evidence 
is reported for A2199 
(Kaastra \etal 1999). These observations are only first steps towards 
assessing the general existence
of this new component in the X-ray spectra of clusters of galaxies. 
The search for nonthermal emission in more clusters is of high importance
as it will allow to derive additional informations on the physical conditions 
of the intracluster medium (ICM) environment, which cannot be obtained by
studying the thermal plasma emission only.

Various interpretations of the HXR emission have been presented since 
its discovery in the Coma cluster spectrum. The most direct explanation is 
inverse Compton (IC) scattering of cosmic microwave background (CMB) photons
by the relativistic electrons responsible of the extended radio emission 
present
in the central region of Coma (Willson 1970). The combined radio
synchrotron and IC HXR fluxes (e.g., Rephaeli 1979) allow to estimate a 
volume-averaged intracluster
magnetic field of $\sim 0.16~\mu G$ (Fusco-Femiano \etal 1999). One 
of the problems with the IC model is that this value of 
the magnetic field in the ICM seems to be at odd with the value determined 
from Faraday rotation of polarized radiation toward the head tail radio galaxy
NGC4869 that gives
a line-of-sight $B\sim 6~\mu G$ (Feretti \etal 1995), and with the
equipartition value in the radio halo, which is $\sim 0.4~h_{50}^{2/7}~\mu G$ 
(Giovannini \etal 1993). We note, however, that Feretti \etal
(1995) also inferred the existence of a weaker and larger scale magnetic field 
component in the range of $0.1-0.2~ h^{1/2}_{50}~\mu G$, and therefore the 
$\sim 6~\mu G$ field could be local. A low average magnetic field is also 
consistent with the model developed by Brunetti \etal (1999), which predicts a
magnetic field strenght decreasing with the distance from the cluster centre.
      
An alternative explanation is nonthermal 
bremsstrahlung (NTB) emission from suprathermal electrons currently accelerated
at energies greater than $\sim$10 keV by shocks or turbulence (Kaastra \etal
1998; Ensslin, Lieu, \& Biermann 1999; Sarazin \& Kempner 1999).
Another and more trivial possibility is that the HXR radiation is due to a hard 
X-ray source present in the external regions of the field of
view of the \bsax PDS (FWHM=$1.3^{\circ}$, hexagonal), as for example a highly
obscured Seyfert 2 galaxy like the Circinus galaxy (Matt \etal 1999). In 
the central region 
($\sim 30^{\prime}$ in radius), the MECS image does not show evidence of this
kind of sources (Fusco-Femiano 1999).  
Hovewer, the detection of a hard nonthermal 
component in other clusters should strongly reduce the probability of 
this last interpretation. 

In this letter we present the results of a long observation of A2256, 
exploiting the unique capabilities of the PDS, onboard \bsax, to search 
for HXR emission (Frontera \etal 1997). The cluster was also observed with
the MECS, an imaging instrument working in the 1.5-10 keV energy range
(Boella \etal 1997).
The galaxy cluster A2256 is similar to the Coma cluster in many 
X-ray properties, as luminosity and presence of substructures. The \rosat 
PSPC observations showed that A2256 is a double X-ray 
cluster (Briel \etal 1991),
 suggesting that a subcluster may be merging with a larger cluster, although 
there is no strong evidence in the temperature map in favour of
an advanced merger (Markevitch \& Vikhlinin 1997), as it is for Coma.
The average gas temperature is $\sim$7 keV, as measured by several X-ray
instruments (David \etal 1993; Hatsukade 1989; Markevitch \& Vikhlinin 1997;
Henriksen 1999).    
Both clusters show a radio halo in the central and periferal regions. 
However, the radio
emission from A2256 is notably complex. The region around the cluster
centre is occupied by an unusual concentration of radio galaxies: at least 
five discrete sources
have been identified with cluster galaxies, but there also two extended 
emission regions which have linear sizes $\leq$1 Mpc   
(Bridle \& Formalont
1976; Bridle \etal 1979; Rottgering \etal 1994).    

Throughout the Letter we assume a Hubble constant of $H_o = 50~km~s^{-1}~Mpc^{-1}~h_{50}$ and
$q_0 = 1/2$, so that an angular 
distance of $1^{\prime}$ corresponds to 92 kpc ($z_{A2256} = 0.0581$; Struble \& Rood 1991). 
Quoted confidence intervals are at $90\%$ level, if not otherwise specified.

\section {PDS and MECS Data Reduction}

The total effective exposure time was
$\sim 1.3\times 10^5$ sec for the MECS and $\sim 7.1\times 10^4$ sec for the 
PDS
in the two observations of February 1998 and February 1999.
The observed count rate for A2256 was 0.497$\pm$0.002 cts/s for the 2
MECS units and 0.27$\pm$0.04 cts/s for the PDS instrument.

Since the source is rather faint in the PDS band (approximately 1.5 mCrab 
in 15-150 keV) a careful check of the background subtraction must be performed. The background sampling
was performed using the default rocking law of the two PDS collimators that 
samples ON, +OFF, ON, -OFF fields for each collimator with a dwell time of 
96" (Frontera et al. 1997). When
one collimator is pointing ON source, the other collimator is pointing 
toward one of the two OFF positions. We used the standard procedure to obtain 
PDS spectra (Dal Fiume et al. 1997), which consists in 
extracting one accumulated spectrum for each unit for each collimator 
position. 
We then checked the two independently accumulated background spectra in the 
two different +/-OFF sky directions, offset by 210' with respect to the 
on-axis pointing direction. 
The comparison between the two accumulated backgrounds 
([+OFF] - [-OFF]) shows a difference with a marginal excess below ~30 keV in
the [+OFF] pointing. This 
excess is much lower
than the signal from the source, but it must not be neglected.
The total excess in the first two equalized energy channels (15-33.5 keV) is 
$0.048\pm 0.024~cts~s^{-1}$, i.e. approximately 2$\sigma$. 
This concentration in only the lowest energy channels implies that the excess
is likely due to contamination by a point source rather than to a statistical
fluctuation.
The total 
source spectrum was 
therefore obtained using only the uncontaminated 
background accumulated pointing at the [-OFF] field. Hovewer, in Section 3
we report the confidence level of the nonthermal emission in excess 
of the thermal one considering the average
of the two background measurements.
The background level of the PDS is the lowest obtained thus far with 
high-energy instruments on board satellites thanks to the equatorial orbit 
and is 
very stable again thanks to the favorable orbit. No modeling of the time
variation of the background is required. 
  
MECS data preparation and linearization was performed using
the {\sc Saxdas} package under {\sc Ftools} environment.
We have extracted a MECS spectrum from a circular region
of 8$^{\prime}$ radius  (corresponding to about 0.8 Mpc) centered on the
primary emission peak. From the ROSAT PSPC radial profile,
we estimate that about 70\% of the total cluster emission falls
within this radius.
The background subtraction has been performed
using spectra extracted from blank sky event files in the same region
of the detector as the source.

A numerical relative normalization factor among the two
instruments has been included in the fitting procedure (see next Section) 
to account for:
a) the fact that the MECS spectrum includes emission out to $\sim$0.8 Mpc from
the X-ray peak, while the PDS field of view (1.3 degrees FWHM) covers the
entire emission from the cluster;
b) the slight mismatch in the absolute flux calibration
of the MECS and PDS response matrices employed
(September 1997 release; Fiore, Guainazzi \& Grandi 1999);
c) the vignetting in the PDS instrument,
(the MECS vignetting is included in the response matrix).
The estimated normalization factor is  $\sim$1.1. In the fitting procedure we
allow this factor to vary within 15\% from the above value to account for
the uncertainty in this parameter.

\section {PDS and MECS Data Analysis and Results}

The spectral analysis of the MECS data alone,
in the energy range 2-9.7 keV and in the central $\sim$0.8 Mpc region,
gives a gas temperature of $kT=7.41\pm 0.23$ keV ($\chi^2$=154.5 for
162 degrees of freedom; hereafter dof),
using an optically thin thermal emission model (MEKAL code on the XSPEC
package), absorbed by a galactic line of sight equivalent hydrogen
column density, $N_H$, of 4.01$\times 10^{20}~cm^{-2}$. This value of
the temperature 
is consistent with the \asca GIS measurement (6.78-7.44
keV; Henriksen 1999), and with the values obtained by previous observations :
the \einstein MPC (6.7-8.1 keV; David \etal 1993) and \ginga (7.32-7.70 keV;
Hatsukade 1989). Also the flux of $\sim 5.3\times 10^{-11}$\erg in the 2-10 keV
energy range is
consistent with the previous measurements. The iron abundance is  
$0.26\pm 0.03$, in agreement with the \asca results
(Markevitch \& Vikhlinin 1997). 

The analysis of the PDS data with a thermal bresstrahlung component gives
a temperature of $\sim$30 keV.   
Fitting the data
 with two thermal components, one of these at the fixed temperature of 7.4 keV,
we obtain a temperature greater than $\sim$90 keV for the second component.   
These unrealistic high values for the gas temperature 
obtained in both the fits are interpreted as a
strong indication that the detected hard excess is due to a nonthermal
mechanism.
  
Figure 1 shows the simultaneous fit to the MECS
and PDS data with a thermal component at the temperature 
of 7.47$\pm$0.35 keV and a normalization factor of $\sim$1.2 for the two 
data sets. The $\chi^2$ is 180.5 for 167 dof.  
Hard X-ray
 radiation at energies
greater than $\sim$20 keV is in excess with respect to the thermal component
at a level of $\sim 4.6\sigma$ and this value is rather stable against 
variation of the
normalization factor. It results slightly lower ($\sim 4.5\sigma$) considering
the average of the two background measurements. Besides, also fitting the 
PDS data alone with a thermal component at the fixed temperature of 7.47 keV
we obtain an excess at a level of 4.3$\sigma$.  
If we introduce a second nonthermal component,  
modeled as a power law,
we obtain the fit shown in figure 2. The $\chi^2$ is 156.6 for 165 dof.  
The improvement with respect to the previous model is significant at more
than the 99.99\% confidence level, according to the F-test.
The confidence contours of the parameters
$kT$ and photon spectral index ($\alpha_X$) show that, at 90\% confidence 
level, the temperature is well determined, 6.8-7.5 keV,
while $\alpha_X$ describes a large interval 0.3-1.7. The presence of 
the nonthermal component has the effect to
slightly decrease the best fit value of the temperature
($6.95^{+0.45}_{-0.35}$ keV),
with respect to the temperature obtained considering only the
MECS data.  The flux of the nonthermal component is rather stable, 
$\sim 1.2\times 10^{-11}$\erg in the 20-80 keV energy range, against 
variations of $\alpha_X$. The contribution
of the nonthermal component to the thermal flux in the 2-10 keV energy range
is $\leq 10\%$ for $\alpha_X\leq$1.70. The analysis of the two
observations with effective exposure times of $\sim$23 ksec (February 1998) and
$\sim$48 ksec (February 1999) for the PDS does not show significant
flux variations. These results and the fact that the two clusters 
with a detected hard X--rays excess (Coma and A2256) both have radio halos, 
strongly support the diffuse nonthermal mechanism as responsible for the 
excess, as discussed in the next section.

\section{Discussion}

A2256 is the second cluster, after Coma (Fusco-Femiano \etal 1999), which
shows hard X-ray radiation up to $\sim$80 keV in the PDS spectrum, with a clear 
excess above the thermal intracluster emission. (A2199 shows only a marginal 
evidence in the external region of the MECS detectors, Kaastra \etal 1999).
We have investigated the possibility that the observed excess in
A2256 could be due by a confusing
source in the field of view of the PDS. The most qualified candidate is
the QSO 4C +79.16
observed by \rosat PSPC with a count rate of $\sim$0.041 c/s (WGA Catalogue). 
With
a typical photon index of 1.8 (\rosat reports a steeper index of $\sim$2.5),
about 1.2 c/s are necessary to account for the observed HXR emission of
$\sim 1.2\times 10^{-11}$\erg in the 20-80 keV energy range of the PDS.
Considering
that the QSO is $\sim 52^{\prime}$ off-axis, an unusual variability
of about two orders of magnitude is required.
There is still the possibility that an obscured
source, like Circinus (Matt \etal 1999), be responsible of the detected
HXR radiation.
Unless the obscured source is within 2$^{\prime}$ of the central 
bright core of A2256,
our analysis of the MECS image excludes the presence of this kind of sources in
the central region ($\sim 30^{\prime}$ in radius) of the cluster.

The application of the inverse Compton model,
based on the scattering of relativistic
electrons with the 3K background photons, appears less straightforward
in A2256 than in the Coma cluster. The radio morphology
is remarkably complex (Bridle \& Fomalont 1976; Bridle \etal 1979;
Rottgering \etal 1994). 
There are at least four radio sources classified as head-tail radio 
galaxies, an ultra steep spectrum source and a diffuse region in the north
with two diffuse arcs ($G$,$H$ according to Bridle \etal 1979), at a
distance of $\sim 8^{\prime}$ from the cluster centre. The extent of 
this diffuse region is
estimated to be 1.0$\times$0.3 Mpc, with a total flux density of 671 mJy at
610 MHz and a rather uniform spectral index of 0.8$\pm$0.1
between 610 and 1415 MHz
(Bridle \etal 1979). The percentage polarization is uniform with an 
average value of 20\%. The
alignement of the electric field vectors suggests a well ordered magnetic field.
The equipartition magnetic field is  1-2$~\mu G$ (Bridle \etal 1979).
A fainter extended emission permeates the
cluster centre (diffuse emission around $D$ in Bridle \etal 1979) 
with a steeper radio spectral index of $\sim$1.8 as estimated by 
Bridle \etal (1979) and in agreement with the 327 MHz data from the
Westerbork Northern Sky Survey (Rengelink \etal 1997).
The total flux density is 100 mJy at 610 MHz and no polarized emission have
been detected from this region. We note that the physical and morphological
properties of the diffuse $D$ emission are consistent with those of central 
halo sources while those in the $G-H$ region are consistent with the properties of
peripheral relic sources as 1253+275 in the Coma cluster.

In addition to the thermal emission, a second component in the X-ray 
spectrum of A2256 was noted by Markevitch \& Vikhlinin (1997) in their
spectral analysis of the \asca data in the central r=$3^{\prime}$ spherical
bin. Although they were not able to firmly establish the origin of this
emission, their best fit is a power law model with photon
index 2.4$\pm$0.3, therefore favouring a nonthermal component. The contribution
of this component to the total flux is not reported in the paper. 
Considering that there are no bright
point sources in the \rosat HRI image, they argued for 
an extended source. Also the joint \asca GIS \& \rxte PCA data analysis
is consistent with the detection of a nonthermal component in addition to the
thermal component. The contribution of this nonthermal component to
the total X-ray flux in the 2-10 energy range is 
$\leq 4\%$. However, a second thermal component (0.75-1.46 keV), instead 
of a nonthermal one, provides a better description of the data 
(Henriksen 1999). The MECS data do not show
evidence of this steep nonthermal component in the central bin of $2^{\prime}$ 
because the energy range is truncated to   
a lower limit of 2 keV (Molendi, De Grandi, \& Fusco-Femiano 2000), while 
a joint fit to the LECS \& MECS data within $4^{\prime}$ does not
show a significant evidence for an additional component at energies lower 
than 2 keV.
   
The power-law component (slope $2.4\pm 0.3$) 
found in 
the analysis of the \asca data 
(Markevitch \& Vikhlinin 1997), and the upper limit of 1.7 for $a_{X}$, 
determined by \bsax data, 
suggest that two tails could be present in the X-ray spectrum of A2256. 
The former might 
due to the central diffuse radio source with the steep index 
$\alpha_R\sim$1.8, and the last to 
the more extended radio emission in the northern region of the cluster with 
the flatter energy spectral index of 0.8$\pm$0.1.  
Assuming that the contribution of the power-law
component, detected by \asca, to the total X-ray flux 
($F_X(2-10 keV)\sim 5\times 10^{-11}$\erg) is $\sim 5\%$, we obtain a 
negligible contribution at PDS energies ($\sim 4\times 10^{-13}$\erg) 
and a magnetic field in the central radio region of $\sim 0.5~\mu G$. 
For the external radio region, with spectral index 0.8, 
the nonthermal X-ray flux   
$f_X(20-80 keV)\simeq 1.2\times 10^{-11}$\erg, derived by the PDS excess, 
leads to a low value of $\sim 0.05~\mu G$. 
Even assuming that a large fraction (say 50$\%$) of the HXR flux is due to 
the several point radio sources
in the central region and/or to the contribution of different
mechanisms, we obtain only a slightly greater value of $\sim 0.08~\mu G$.
The combined fit of \asca GIS and \rxte PCA data (Henriksen 1999) gives an
upper limit of $2.64\times 10^{-12}$\erg in the 2-10 keV energy range 
 for the nonthermal component that
corresponds to a lower limit for the volume-averaged intracluster magnetic
field, $B$, of 0.36$~\mu G$
($\alpha_R=1.8$). Considering that the HXR flux detected by the PDS is
in agreement with the above value, we would obtain  
a value for $B$ consistent with that derived by the
GIS \& PCA data, but the fit to the MECS \& PDS
data is
unacceptable for $\alpha_X = 1+\alpha_R = 2.8$.

The previous scenario of a decreasing intracluster magnetic field from the
cluster center would be difficult to reconcile with the stronger periferal
radio region and higher equipartition magnetic field with respect to
the central radio halo. Therefore, we could consider the possibility, 
recently suggested by Brunetti \etal (1999), that
the HXR IC spectrum may be flatter than the synchrotron radio spectrum because
of the acceleration and energy loss processes that produce an electron spectrum
with different slopes. A different electron spectrum index for HXR and radio
emissions is more likely for low magnetic fields which require higher electron 
energies
for synchrotron than for IC radiation. This could explain the better fit
to the PDS data of A2256 with $\alpha_X < 1+\alpha_R$=1.8. Besides, this model
suggests an alternative interpretation of the HXR excess of A2256. 
We can consider that a single hard tail is present in the X-ray spectrum
of the cluster with index $\alpha_X\leq$1.7, as detected by the PDS. The 
electron spectrum responsible of this HXR IC emission can produce radio 
emission with spectral index $\alpha_R > \alpha_X$-1=0.7
with a resulting mean volume-averaged intracluster
magnetic field higher than the one we derive from the classical IC model.
 
A different  mechanism which may produce HXR radiation is given by
nonthermal bremsstrahlung. 
Sarazin \& Kempner (1999) suggest that all or part of the HXR emission 
detected in the Coma
cluster might be NTB from suprathermal
electrons formed through current acceleration of the thermal gas,
either by shocks or turbulence in the ICM. For A2256 the MECS \& PDS 
measurements determine a power-law
momentum spectrum of the electrons with index $\leq 2\alpha_X-1=2.4$ (90\%). 
The
consequence is that an accelerating electron model with flat spectrum
produce more IC HXR emission than the NTB mechanism, unless the electron
spectrum cuts-off or steepens at high energies. Besides, these models 
produce more radio emission than observed if $B$ is $\geq 1~\mu G$.

\acknowledgments 

We thank P. Giommi for useful suggestions regarding data
analysis, G. Brunetti for discussions on the interpretation of the results,
 and the referee for valuable comments.

\newpage

\figcaption[mecs_pds.eps]{MECS and PDS data. The continuous line represents 
a thermal component at the average cluster gas temperature of 
7.47$\pm$0.35 keV. The errors bars are quoted at 1$\sigma$ level.}

\figcaption[a2256_fig2.eps]{The continuous line is the best-fit to the 
MECS (2-9.7 keV) and PDS (15-150 keV) data. The dashed line represents 
the thermal component (kT=6.95$^{+0.45}_{-0.35}$ keV), while the dot-dashed 
line is 
the non-thermal component with a spectral index of 1.4. A value of
1.21 takes into account the relative normalization of the two instruments. The
 reduced $\chi^2_{red}$ is 0.95 for 165 degrees of freedom.} 

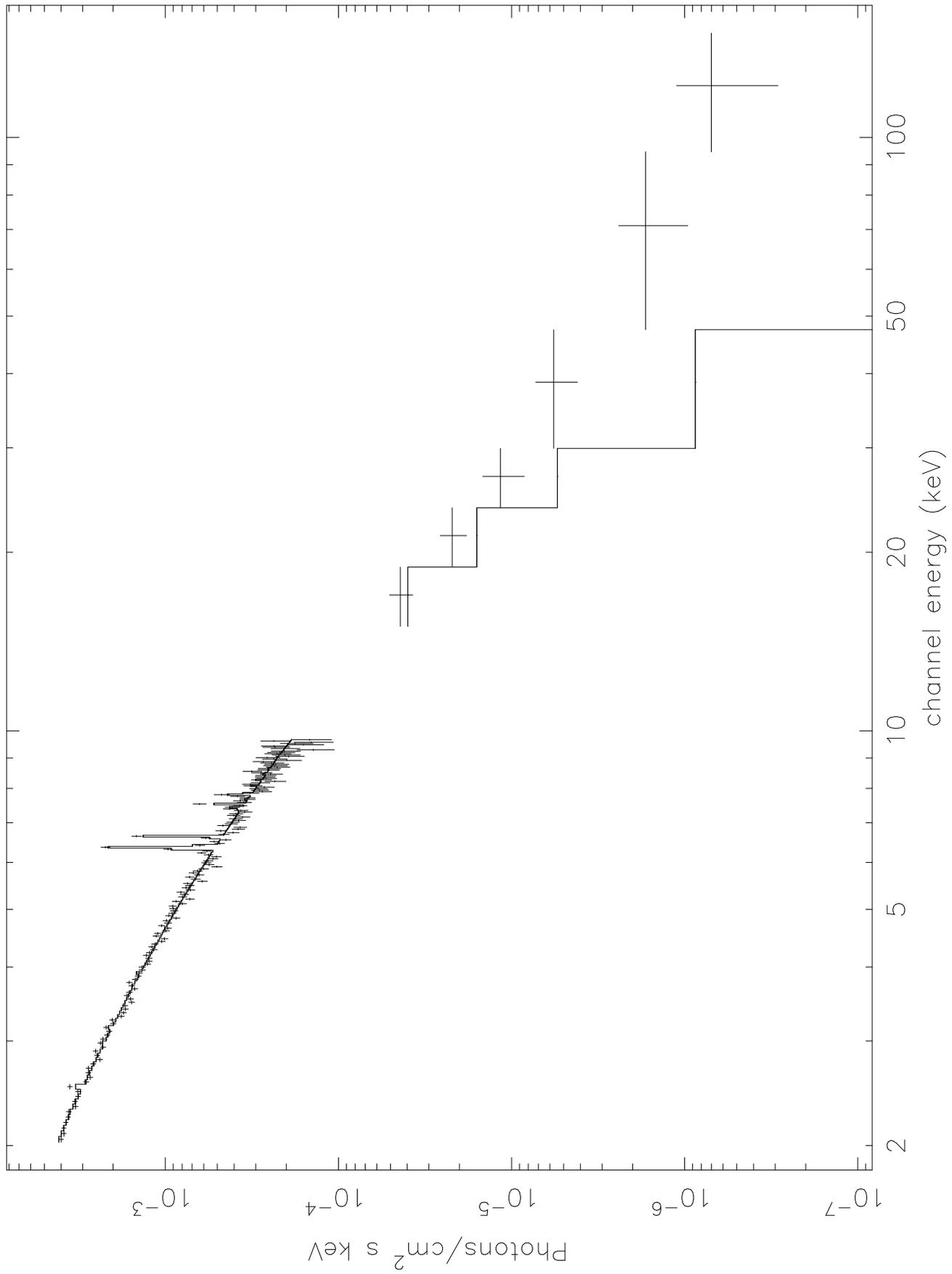
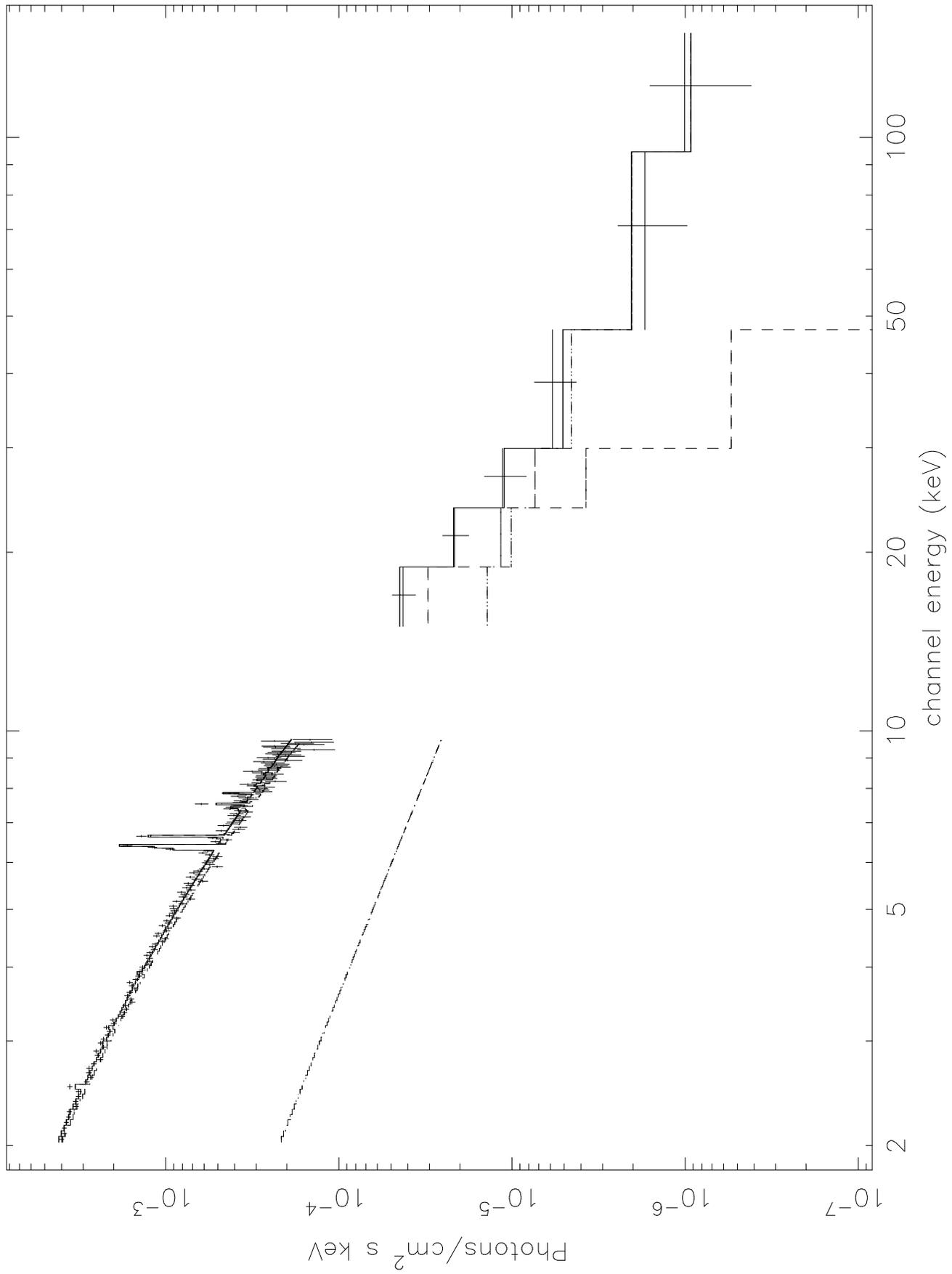
\end{document}